\documentclass[CRPHYS,Unicode,manuscript]{cedram}
\usepackage{bm}
\usepackage{braket}
\usepackage{booktabs}

\title{Connected correlations in cold atom experiments}

\newcommand{\LCF}{Laboratoire Charles Fabry, Institut d'Optique Graduate School, CNRS, Université Paris-Saclay, 91127 Palaiseau, France}

\author{\firstname{Thomas} \lastname{Chalopin}\CDRorcid{0000-0001-7633-0442}}
\addressSameAs{1}{\LCF}
\email[T. Chalopin]{thomas.chalopin@institutoptique.fr}

\author{\firstname{Igor} \lastname{Ferrier-Barbut}\CDRorcid{0000-0002-4707-0474}}
\addressSameAs{1}{\LCF}
\email[I. Ferrier-Barbut]{igor.ferrier-barbut@institutoptique.fr}

\author{\firstname{Thierry} \lastname{Lahaye}\CDRorcid{0000-0002-2260-9859}}
\addressSameAs{1}{\LCF}
\email[T. Lahaye]{thierry.lahaye@institutoptique.fr}

\author{\firstname{Antoine} \lastname{Browaeys}\CDRorcid{0000-0001-9941-8869}}
\addressSameAs{1}{\LCF}
\email[A. Browaeys]{antoine.browaeys@institutoptique.fr}

\author{\firstname{David} \lastname{Clément}\CDRorcid{0000-0003-1451-0610}\IsCorresp}
\address{\LCF}
\email[D. Clément]{david.clement@institutoptique.fr}


\begin{abstract}
The recent development of single-atom-resolved probes has made full counting statistics measurements accessible in quantum gas experiments.
This capability provides access to high-order moments of physical observables, from which cumulants, or equivalently connected correlations, can be precisely determined.
Through a selection of recent cold atom experiments, this article illustrates the significance of connected correlations in characterizing ensembles of interacting quantum particles.
First, non-zero connected correlations of order $n>2$ unambiguously identify non-Gaussian quantum states.
Second, connected correlations of order $n$ identify clusters made of $n$ elements whose statistical properties are irreducible to combinations of smaller clusters.
The ability to identify such multi-particle clusters offers a an interesting perspective on
strongly correlated quantum states of matter at the microscopic scale. 
\end{abstract}

\begin{altabstract}
Le développement récent de méthodes expérimentales permettant de détecter les atomes un par un a rendu possible la mesure des statistiques de comptage dans les expériences de gaz quantiques.
Cette capacité permet d'accéder aux moments d'ordre élévé d'observables physiques, à partir desquels les cumulants, ou de manière équivalente les corrélations connectées, peuvent être précisément déterminés.
A travers une sélection d'expériences récentes utilisant des atomes froids, cet article illustre l'importance des corrélations connectées pour caractériser des ensembles de particules quantiques en interaction.
Premièrement, des corrélations connectées non nulles d'ordre $n>2$ identifient sans ambiguïté des états quantiques non gaussiens.
Deuxièmement, les corrélations connectées d'ordre $n$ identifient des sous-ensembles composés de $n$ éléments dont les propriétés statistiques ne peuvent pas être comprises par des combinaisons d'ensembles plus petites.
La capacité d'identifier de tels ensembles à $n$-corps offre un point de vue intéressant sur la matière fortement corrélée à l'échelle microscopique.
\end{altabstract}

\begin{document}

\maketitle

\newpage
\tableofcontents

\section{Introduction}

Understanding many-body quantum physics remains one of the most formidable challenges in physics \cite{morosan:2012, georgescu:2014, langen:2015a, monroe:2021, grass:2025}.
While \emph{ab-initio} methods offer a rigorous approach to describing quantum many-body systems from first principles, they are often computationally prohibitive and can obscure the intuitive understanding of the microscopic processes that give rise to macroscopic phenomena.
Traditionally, the probing of strongly correlated electronic materials has heavily relied on measuring response functions \cite{basov:2011}.
Specifically, one-particle spectral functions, obtained through ARPES experiments \cite{damascelli:2003, sobota:2021}, and dynamical structure factors, derived from neutron scattering \cite{muhlbauer:2019}, have been indispensable.
These methods effectively probe one-particle (Green functions) and density correlations, offering direct insights into one- and two-body correlations within quantum systems such as superconductors and frustrated magnets, as well as quantum gas experiments \cite{dao:2007, stewart:2008,veeravalli:2008, clement:2009, hart:2015, shao:2024}.
However, in strongly correlated matter where higher-order correlations ($n$-body correlations with $n>2$) are prevalent, traditional spectral functions serve only as indirect probes, with their shapes and dependence on experimental parameters influenced by higher-order correlations. Extracting the contribution of specific high-order correlations from their spectra often remains a challenge.

Recent advances in detection techniques of synthetic quantum systems of ions \cite{monroe:2013}, atoms \cite{ott:2016a, gross:2021} and superconducting qubits \cite{devoret:2013, kjaergaard:2020} have opened new avenues for exploring many-body physics and correlations in experiments through single-particle-resolved probes and measurements of full counting statistics (FCS)~\cite{bohnet:2016, schweigler:2017, rispoli:2019, makhalov:2019, ebadi:2021, chen:2023, herce:2023, joshi:2025}.
Assuming that sufficient statistics is experimentally accessible, these methods offer direct access to higher-order moments of physical observables. Therefore, they provide a means for directly investigating the intricate $n$-body correlations that define much of strongly correlated condensed matter.
By probing individual quantum constituents --- such as electrons, atoms, or spins --- one may hope to reveal the mechanisms that govern collective quantum behaviour, offering new insights into this complex and fundamental question with respect to other types of measurements.

Measuring high-order moments to identify the contribution of \emph{connected} sets of particles was proposed long ago, beginning with the work of Ursell \cite{ursell:1927}. 
In this context, a connected set of $n$ particles designates a group of $n$ particles all correlated with one another, {\it i.e.} a group that cannot be described as formed by two (or more) independent sub-groups.
In his efforts to better understand classical gases, Ursell aimed at efficiently evaluating Gibbs' phase integral by focusing on specific groupings of molecules --- those forming connected clusters.
To this aim, he exploited the concept of cumulants of a set of $N$ random variables $\{ X_{i} \}_{N}$, now referred to as the Ursell function $u(X_{1}, X_{2}, ... , X_{N})$, or the \emph{connected correlation functions} $\braket{X_1X_2\cdots X_N}_{\rm c}$.
Cumulants isolate genuine correlations: the $n$-th order cumulant $\kappa_{n}$ captures correlations that cannot be explained by lower-order correlations.
The key idea of Ursell was to determine a finite (and ideally minimal) number of such cumulants (or connected correlations) that could provide an effective description of the system. 
The Bogoliubov-Born-Green-Kirkwood-Yvon (BBGKY) hierarchy in classical statistical physics is another example of a similar approach \cite{reichl:1998}. 
Identifying a hierarchy in the cumulant contributions was then proven unattainable for strongly correlated bosons \cite{robinson:1965}, but we are not aware of such a conclusion for fermions or spin systems.
Hence, an expansion in orders of cumulants does not necessarily provide a viable description of the many-body state. 
Nevertheless, cumulants offer a clean, intuitive and physically meaningful way to characterise fluctuations experimentally, especially when dealing with complex or strongly interacting systems.

The present article illustrates two key aspects of cumulants that are  useful in the context of strongly correlated quantum matter. 
A first crucial feature of cumulants or connected correlations is their capability to identify non-Gaussian behaviour: any non-zero cumulant $\kappa_{n} \neq 0$ of order $n>2$ indeed reveals the presence of non-Gaussian statistics. 
In addition to their capability to distinguish Gaussian from non-Gaussian statistics, non-zero cumulants reveal genuine forms of clustering which can help building a physical picture of the system. 
Their measurements in experiments exploring quantum many-body phases and dynamics has attracted growing interest over the past years. 
This capability spans a wide range of platforms, including photonic systems, superconducting qubits, trapped ions, and ultracold atoms. 
This work focuses on recent experiments realised with cold atom systems. 
It aims to offer a brief, non-exhaustive overview of how connected correlations are measured and employed to uncover both Gaussian and non-Gaussian aspects of quantum states, characterise strongly correlated matter or highlight the relevant features of quasiparticles emerging in interacting systems. 
Following a short introduction to the concept of cumulants, we highlight their applications in systems of interacting bosons, fermions and spins.

\section{Cumulants or connected correlations}

\subsection{Definition of cumulants}

\paragraph{Cumulants -- connected correlations.} 
The cumulants, or connected correlations, $\kappa_n=\braket{X^n}_{\rm c}$ of a random variable $X$ are defined from their generating function $K_X(t)=\log [\braket{e^{tX}}] = \sum_n \kappa_n t^n / n!$.
An example of such a generating function is the free-energy $F$ in statistical physics, defined as $F(\beta)=-1/\beta \ \log (\braket{e^{-\beta E}})$ \cite{huang:1987}, in the canonical ensemble: $F$ is the generating function of the cumulants $\kappa_n$ of the energy $E$. 
The moments $m_{n}=\braket{X^n}$ are related to the cumulants $\kappa_n=\braket{X^n}_{\rm c}$ through the relation,
\begin{equation}
    m_{n}=\kappa_{n} + \sum_{j=1}^{n-1} \  \frac{(n-1)!}{j!(n-1-j)!} \ m_{j} \kappa_{n-j}.
\end{equation}
A similar relation relates cumulants to moments.

Of particular interest to the present work is a modified version of the above relation, where the $n^{\rm th}$-order cumulant $\kappa_n$ is expressed as a function of the $n^{\rm th}$-order moment $m_n$ and cumulants of order $n'<n$,
\begin{equation}
\kappa_n = m_n - \sum_{P_{j} ; j=1}^{n-1} \ \prod_{B \in P_{j }} \kappa_{|b|}
\label{Eq:Kn-recursive}
\end{equation}
where $P_{j}$ are all the partitions of $n$ objects into lists $B$ of sub-ensembles whose largest sub-ensemble contains $j$ objects ($|b|$ is the size of the sub-ensembles in the list $B$). This formula expresses a simple and useful picture of the cumulants \cite{novak:2014}: the $n^{\rm th}$-order cumulant counts the number of connected graphs in a vertex set with $n$ elements. 
This number is equal to the total number of possible graphs $m_n$, to which the contributions of connected graphs of order $n'<n$ are subtracted. Therefore, the connected correlations of order $n$ identify the number of connected sets of exactly $n$ correlated elements.

\begin{figure}[!ht]
\includegraphics[width=\columnwidth]{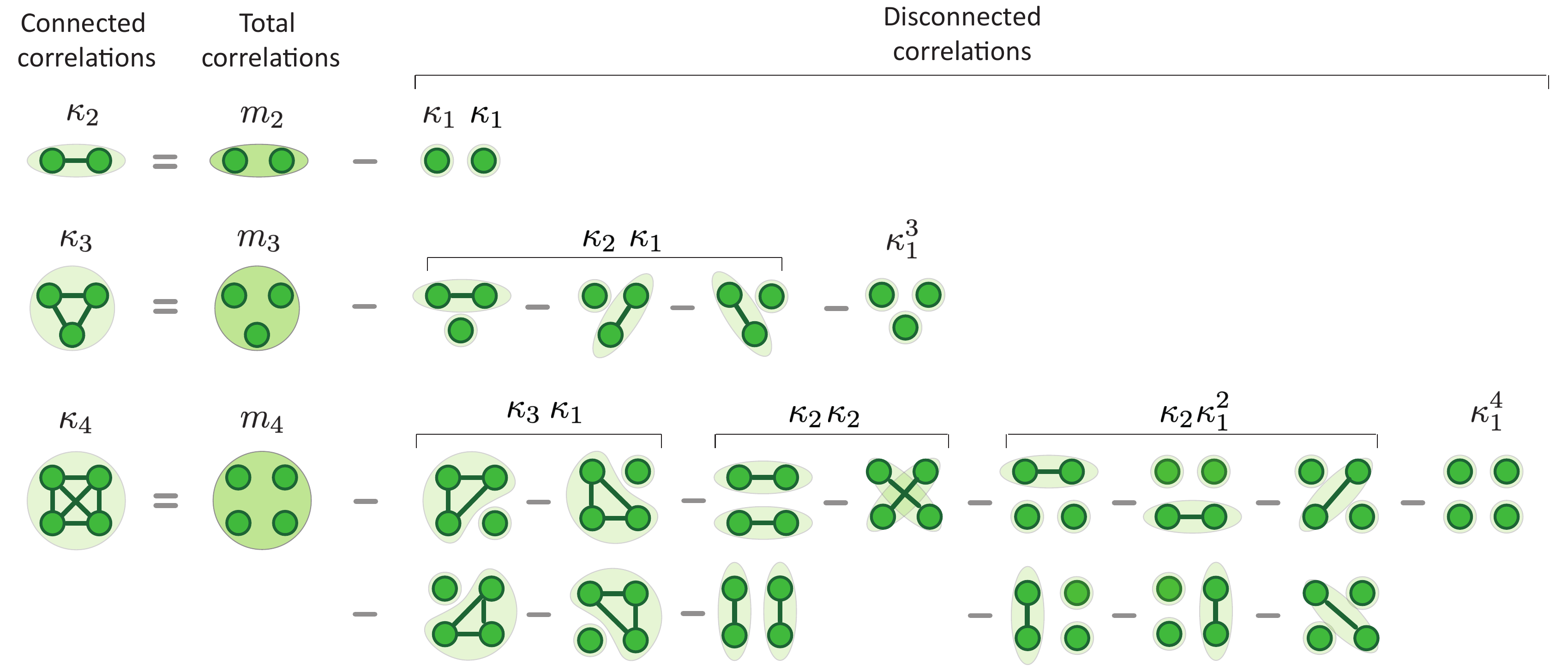}
\caption{Illustration of the decomposition of cumulants of order $n$ as the difference between the total correlations and all the contributions from cumulants of order $n'<n$ \cite{rispoli:2019}.}
\label{fig1}
\end{figure}

As illustrated in Fig.~\ref{fig1}, it follows that the connected correlations can be expressed from subtracting from the (total) correlations $m_n$ of order $n$ the contributions from all connected correlations of order $n' < n$. The difference between the total correlations and the connected correlations is referred to as the disconnected correlations (note that disconnected correlations of order $n$ include connected correlations of order $n-1$, $n-2$, ...). To complete this illustration, we write the connected correlations of lowest order,
\begin{eqnarray}
\braket{ X_{i} }_{\rm c} &=& \braket{ X_{i} } \nonumber \\ 
\braket{ X_{i} X_{j} }_{\rm c} &=& \braket{ X_{i} X_{j} } - \braket{ X_{i} }_{\rm c} \braket{ X_{j} }_{\rm c} \nonumber \\
\braket{ X_{i} X_{j} X_{k}}_{\rm c} &=& \braket{ X_{i} X_{j} X_{k}} - \left ( \braket{ X_{i} }_{\rm c} \braket{ X_{j} X_{k}}_{\rm c} + \braket{ X_{j} }_{\rm c} \braket{ X_{i} X_{k} }_{\rm c} + \braket{ X_{k} }_{\rm c} \braket{ X_{i} X_{j}}_{\rm c} \right ) - \braket{ X_{i} }_{\rm c} \braket{ X_{j} }_{\rm c} \braket{ X_{j} }_{\rm c}\nonumber \\
\end{eqnarray}

The cumulants of low order have names: mean, variance, skewness and kurtosis for $\kappa_1$, $\kappa_2$, $\kappa_3$ and $\kappa_4$ respectively. While the first three cumulants are identical to the centered moments of corresponding order, $\kappa_n = \braket{ (X-\braket{ X})^n } $ for $n=1, 2, 3$, this is not the case anymore from $n>3$, \emph{e.g.} $\kappa_4 = \braket{ (X-\braket{ X})^4 } - 3 \kappa_2^2$.

\paragraph{Cumulants of classical Gaussian random variables.}
A Gaussian random variable is a random variable whose connected correlations (or cumulants) are zero at any order larger than 2, {\it i.e.} $\kappa_{n}=0$ for $n>2$. This definition is equivalent to stating that the knowledge of the mean and standard deviation fully defines a Gaussian random variable.
Otherwise stated, connected sets with more than 2 particles are not present in systems with Gaussian statistics.

\subsection{Cumulants of quantum observables}

\paragraph{Connected correlations of quantum states.} 
In the quantum mechanical description, a system is described by a density operator $\hat{\rho}$ and the statistical average $\braket{ \cdots }$ of an observable $\hat{A}$ is $\braket{ \hat{A} } = \textrm{Tr}[\hat{\rho} \hat{A}]$.
The moments and the cumulants (or connected correlations) of $\hat{A}$ have the same definition as in the previous paragraph. 
The cumulant generating function is defined as $K_{\hat{A}}(\lambda)=\log [\braket{e^{\lambda \hat{A}}}]$ and cumulants are obtained from its derivatives at $\lambda=0$. 
In contrast to the classical case, however, care must be taken in these definitions as the observable $\hat{A}$ and the Hamiltonian $\hat{H}$ of the system may not commute, with the consequence that different definitions lead to different results. 
For instance, at equilibrium where $\hat{\rho}=e^{-\beta \hat{H}}/Z$, one may find an alternative, symmetrised definition of the cumulant generating function to the one given here, namely $K_{\hat{A}}(\lambda)=\log \textrm{Tr}[e^{\lambda \hat{A}/2} \hat{\rho} e^{\lambda \hat{A}/2}]$. 
When $\hat{A}$ and $\hat{H}$ do not commute, this alternative definition leads to different values for the cumulants. 
Similar attention to normal ordering must be paid to the case of an observable $\hat{A}$ involving non-commuting operators. 
Its cumulants are obtained from the generating function of the joint cumulants associated to the operators in normal order \cite{carmichael:2008}.

\paragraph{Link to normalised correlation functions.} 
In the field of quantum optics, R.\,Glauber introduced the normalised correlation functions for a light field associated with photon  operators \cite{glauber:1963}. For any creation and annihilation $\hat a^\dag$ and $\hat a$ (both bosonic and fermionic operators), one can define a normalised correlation function as
\begin{equation}
g^{(n)}\left ( \bm x_1, \ldots, \bm x_n  \right ) = \frac{\braket{ \hat a^{\dagger}(\bm x_1) \cdots  \hat a^{\dagger}(\bm x_n) \hat a(\bm x_n)\cdots \hat a(\bm x_1)}}{\braket{ \hat I(\bm x_1) } \braket{ \hat I(\bm x_2) } \cdots \braket{ \hat I(\bm x_n) }} = \frac{G^{(n)}\left ( \bm x_1, \ldots, \bm x_n  \right )}{\braket{ \hat I(\bm x_1) } \braket{ \hat I(\bm x_2) } \cdots \braket{ \hat I(\bm x_n) }},
\end{equation}
with $\hat I(\bm x_j) = \hat a^{\dagger}(\bm x_j) \hat a(\bm x_j)$ the particle number operator at coordinates $\bm x_j = (t_j, \bm r_j)$.
Note that the ordering of the creation and annihilation operators plays an important role in the case of fermionic operators. 
The normalised correlation function $g^{(n)}$ reflects the total correlations and it can be related to the connected correlations using Eq.~(\ref{Eq:Kn-recursive}).
For instance, the connected ``intensity'' correlations relates to the normalised two-body correlation function as follows 
\begin{eqnarray}
\braket{ \hat I(\bm x_1) \hat I(\bm x_2) }_{\rm c} 
&=& \left [ g^{(2)}(\bm x_1,\bm x_2) - 1 \right ]  \ \braket{ \hat I(\bm x_1) } \braket{ \hat I(\bm x_2) } + 
\delta(\bm x_1 - \bm x_2) \braket{ \hat I(\bm x_1) },
\end{eqnarray}
where $\delta(\bm x_1 - \bm x_2)$ is the Dirac function which is non-zero only when $\bm x_1 = \bm x_2$. 
In terms of field operators, the normalised two-body correlation function $g^{(2)}$ corresponds to a fourth-order cumulant and contains many contributions (see Fig.~\ref{fig1}),
\begin{equation}
    g^{(2)}(\bm x_1,\bm x_2) \propto \braket{\hat a^{\dagger}(\bm x_1)  \hat a^{\dagger}(\bm x_2) \hat a(\bm x_2) \hat a(\bm x_1)}_{\rm c} + \braket{\hat a^{\dagger}(\bm x_1)  \hat a^{\dagger}(\bm x_2) \hat a(\bm x_2)}_{\rm c} \braket{ \hat a(\bm x_1)} + ...
\end{equation}
We emphasise that $g^{(2)}$ contains the \emph{total} correlations (see Fig.~\ref{fig1}), and  is an intensive quantity due to its normalisation. 
For example, in quantum optics, $G^{(2)}$ increases with the intensity, {\it i.e.} the number of photons in a temporal mode, while $g^{(2)}$ does not.
This contrasts with \emph{connected} correlations $\braket{\cdots}_{\rm c}$ that effectively count the number of connected sets and thus grow with system size.
In that regard, connected intensity (or number) correlations can easily be interpreted as the difference between correlated pairs compared to the situation of independent random variables.

\subsection{Motivations for measuring cumulants}

In the present article, we illustrate on a few examples from cold atoms experiments why cumulants are interesting. 
We have identified two main motivations for measuring cumulants: revealing non-Gaussian quantum states and identifying correlated clusters. 
In the context of simulating and understanding strongly correlated quantum matter, we believe that these two motivations are extremely fruitful for experiments and represent strong assets associated with the measurements of high-order correlations. 

\paragraph{Distinguishing Gaussian from non-Gaussian quantum states.}  
A Gaussian quantum state is a state described by a Gaussian Wigner function, or a Gaussian density matrix \cite{gardiner:2004}. 
Similarly to the case of a classical Gaussian variable mentioned previously, any connected correlations involving $n \geq 3$ creation/annihilation operators are zero. 
This implies that a measurement of a non-zero high-order ($n\geq3$) cumulant of creation/annihilation operators unambiguously indicates that the quantum state is non-Gaussian.

A parallel can somewhat be drawn with the theoretical description of quantum systems of interacting particles, as illustrated by  table \ref{TableCumTheo}.
Mean-field theories do not account for quantum fluctuations, establishing equations for one particle in a (classical) field.
Therefore they do not include any connected sets of modes\footnote{A mode is a quantized excitation of a field, associated to a creation/annihilation operator. For instance, a mode in momentum space is associated to the creation operator $\hat{a}_k^{\dagger}$ of having a excitation with momentum $k$. The BCS and Bogoliubov theories predict pairing of modes at opposite momenta $k/-k$ (see below).}. 
Gaussian theories --- such as Bogoliubov theory, Bardeen-Cooper-Schrieffer (BCS) theory and Luttinger liquid theory --- include linearised quantum fluctuations that give rise to correlations between two modes. 
As a result, they can be viewed as approaches considering the contributions of connected sets with up to two modes. 
Beyond the linearisation of quantum fluctuations and beyond Gaussian theories, descriptions of strongly correlated quantum matter include the contributions of connected clusters with more than two modes.

\begin{table}[h]
\begin{center}
\setlength{\tabcolsep}{.25cm}
\renewcommand{\arraystretch}{1.3}
\begin{tabular}{p{2.2cm} p{4.2cm} p{6cm}
}
\toprule
\textbf{Cumulants of operators} & \textbf{Theoretical description} & \textbf{Type of correlations}  \\
\midrule \midrule
$\kappa_{n}=0$ for $n>1$ & Mean-field theories & No correlations between modes 
\newline Single-mode correlations induced by quantum statistics\\
$\kappa_{n}=0$ for $n>2$  &  Gaussian theories \newline Free quasi-particles theories  & Gaussian correlations coupling two modes \\
$\kappa_{n}\neq0$ for $n>2$  & Theories of strongly correlated states & Non-Gaussian correlations between many modes\\
\bottomrule
\end{tabular}
\end{center}
\caption{Cumulants of operators and their relation to theoretical descriptions of quantum states with increasing number of correlated modes.}
\label{TableCumTheo}
\end{table}

From an experimental point of view, correlations between modes ({\it i.e.} field correlations) are routinely probed through homodyne and heterodyne interferometers in optics.
In contrast, correlations between modes of massive particles are rather revealed indirectly from measuring correlations between particle numbers. 
This difference between mode correlations and particle correlations has important consequences when identifying the properties of quantum states. 
For instance, mode entanglement and particle entanglement are not the same \cite{horodecki:2009, morris:2020}. 
Moreover, when attempting to reveal the non-Gaussian nature of a quantum state --- defined from a non-zero operator cumulant of order $n>2$ --- measuring two-particle correlations can be sufficient, as it indirectly provides information about 4-mode correlations (see section~\Ref{Sec-NonGauss-Bosons}).

\paragraph{Identifying correlated clusters.} One attempt in understanding correlated quantum matter is to identify the presence of correlated clusters of modes (or particles) at the microscopic level. 
Celebrated examples are pairing mechanisms at the heart of superconductivity --- such as described by Bardeen-Cooper-Schriefer theory for instance --- or superfluidity in liquid Helium-3. 
Larger correlated clusters involving more than two modes (or particles) are also present in strongly correlated matter, as we shall illustrate below. 
In experiments, measuring $n$-body correlations is not sufficient to reveal these microscopic $n$-uplet clusters --- pairs, triplets, quadruplets, ... --- since $n$-body correlations also include contributions from $n'$-uplets with $n'<n$ (see Fig.~\ref{fig1}). 
Instead, $n$-order cumulants or connected correlations are the adequate quantities to study as they contain only the genuine contribution from $n$-uplets. This is illustrated through the fact that (total) $n$-body correlations $G^{(n)}$ can take non-zero values even in the absence of $n$-uplet clusters, {\it  i.e.} when $n$-body connected correlations $G_{\rm c}^{(n)}$ are equal to zero. 

\paragraph{Continuous phase transitions}

The description of phase transitions in statistical physics relies on the scale invariance of the system's fluctuations close to the transition \cite{nishimori:2010}. 
A well-known consequence of scale invariance is the emergence of algebraic scaling laws for physical quantities, with critical exponents determined by the universality class of the phase transition. 
This prediction has found experimental confirmation in a large variety of systems ranging from condensed matter to ultracold gases.

Another property deriving from scale invariance is that the statistics of the order parameter assume a universal scaling function. 
Notably, this universal distribution is expected to be non-Gaussian near the critical point, where fluctuations occur across all spatial scales and the Central Limit Theorem no longer applies \cite{bouchaud:1990, Balog:2022}.
Higher-order cumulants of the order parameter are particularly well-suited to uncover the non-Gaussian nature of these fluctuations. 
However, due to the challenges in measuring high-order cumulants --- or equivalently the full probability distribution --- of an order parameter, experimental confirmation of its non-Gaussian behaviour at the phase transition remains limited \cite{joubaud:2008}.

Recently, the work of Allemand \emph{et al.} with ultracold lattice bosons exploited high-order cumulants of the condensate order parameter to identify the non-Gaussian critical regime of the superfluid-to-Mott phase transition \cite{allemand:2025}. 
In addition, it uncovered universal oscillations of high-order cumulants across phase transitions occurring in finite-size systems.
Interestingly, a similar approach could be applied to study various phase transitions, belonging to different universality classes, as several atomic platforms are capable of probing the statistics of order parameters \cite{makhalov:2019,ebadi:2021,chen:2023,ho:2025}.

\section{Models of interacting bosons and fermions}

\subsection{Quantum statistics: single-mode correlations}

In the absence of interactions, the unique origin of correlations is quantum statistics. The latter take place within a single mode, resulting in single-mode correlations only. They manifest as a bunching effect in the case of bosons and an anti-bunching effect in the case of fermions. While these correlations are not the ones we are mostly interested to discuss in this article, we briefly comment on them, as well as on the Siegert relation and Wick's theorem \cite{cohen-tannoudji:2001}.

Isserlis' theorem or Wick's theorem relates the moments $m_{n}$ of a Gaussian variable, or of a Gaussian field, to a combination of moments of order $n' \leq 2$.
Such a relation can be derived from the connected correlations: using Eq.~\eqref{Eq:Kn-recursive} and the hypothesis that the statistics is Gaussian, \emph{i.e.} that any connected correlations of order $n \geq 3$ vanish.
Consider for instance two bosonic modes labeled 1 and 2 respectively, with their respective annihilation operators $ \hat a_1$ and $\hat a_2$. The Gaussian nature of the state implies that $\braket{ \hat a^{\dagger}_1 \hat a^{\dagger}_2 \hat a_2 \hat a_1 }_{\rm c} =0$, from which one obtains $\braket{ \hat a^{\dagger}_1 \hat a^{\dagger}_2 \hat a_2 \hat a_1 } =  D_{4}^3 + D_{4}^{2} + D_{4}^{1}$ where $D_4^{j}$ corresponds to the disconnected correlation of order 4 involving connected (operator) correlations of order $j$ (see Fig.~\ref{fig1}).
In the case of an incoherent field, \emph{i.e.} $\braket{ \hat a_i } =0$, one has $D_{4}^3=D_{4}^1=0$ and $D_{4}^2= |\braket{ \hat a^\dagger_{1} \hat a_{2} } |^2 + \braket{ \hat a^{\dagger}_{1} \hat a^{\dagger}_{2} } \braket{ \hat a_{1} \hat a_{2} } + \braket{ \hat{n}_{1}} \braket{ \hat{n}_2}$.
With the additional assumption $\braket{ \hat a_{1} \hat a_{2} }=0$ --- for instance valid in systems with chaotic statistics or with particle number conservation ---, one is left with 
\begin{equation}
\braket{ \hat a^{\dagger}_1 \hat a^{\dagger}_2 \hat a_2 \hat a_1 } = \braket{ \hat a^{\dagger}_1 a_{1} } \braket{ a^{\dagger}_2 \hat a_{2}} + |\braket{ \hat a^\dagger_{1} \hat a_{2} } |^2.
\label{Eq:thermal}
\end{equation}
This equation is well-known in the context of quantum optics and referred to as the Siegert relation when a single mode is probed.
Considering the light emitted by an incoherent source (for which $\braket{\hat a(t)}=\braket{\hat a(t)^2}=0$ with $\hat a(t)$ being a stationary process), it reads \begin{equation}
g^{(2)}(\delta t) = 1 + | g^{(1)}(\delta t) |^2
\end{equation}
where $g^{(1)}(\delta t) = \braket{\hat a^{\dagger}(t) \hat a(t+\delta t)}/ \sqrt{\braket{\hat a(t)^{\dagger} \hat a(t) } \braket{\hat a(t+\delta t)^{\dagger} \hat a(t+\delta t) }}$ characterises the first-order temporal coherence of the source.
The above derivation simply relies on the assumption that the light field has Gaussian statistics and zero mean.
Chaotic or thermal states, which are often encountered as they result from the incoherent sum of many contributions and application of the Central Limit Theorem \cite{goodman:2015}, share those properties and, in turn, follow the Siegert relation.

When considering a single Gaussian bosonic mode, a similar approach exploiting the structure of connected correlations (equivalent to Wick's theorem) can be applied in the presence of a finite coherence of the field. 
Using the vanishing of high-order cumulants, {\it e.g.} $\braket{\hat a^{\dag}\hat a^\dag \hat a }_{\rm c} = 0$ and $\braket{\hat a^{\dag}\hat a^\dag \hat a \hat a}_{\rm c} = 0$, a little bit of algebra \cite{kubo:1962} yields the generic expression,
\begin{equation*}
\braket{\hat a^\dag \hat a^\dag \hat a \hat a} = |\braket{\hat a^2}|^{2} + 2\braket{\hat a^\dag \hat a}^2 - 2|\braket{\hat a}|^{4}
\end{equation*}
This expression generalises the Siegert relation (see Eq.~\eqref{Eq:thermal}) in the case of a single bosonic mode. 
More specifically, it contains terms with anomalous averages that affect the statistical properties of the state. For a bosonic mode with thermal statistics, one has $\braket{\hat a}=\braket{\hat a^2}=0$ and one recovers $\braket{\hat a^{\dagger}\hat a^{\dagger}\hat a\hat a} = 2\braket{a^\dag a}^2$, {\it i.e.} a perfectly-contrasted bunching. 
In the case of a coherent state $\ket{\alpha}$ --- which is a Gaussian state obtained from the vacuum state $\ket{0}$ by the displacement operator $\hat D(\alpha)$, $\ket{\alpha}=D(\alpha) \ket{0}$ --- anomalous averages are non-zero. 
From the properties of the displacement operator, one finds $\braket{\hat a} = \alpha$  and $\braket{\hat a^2} = \alpha^2$, as well as $\braket{\hat a^\dag a} = |\alpha|^2$, leading to $\braket{\hat a^{\dagger}\hat a^{\dagger}\hat a\hat a} = |\alpha|^4 = \braket{\hat a^\dag \hat a}^2$ signaling the absence of bunching. 
This result is straightforwardly generalised to higher-order correlations in the same mode $\{ {\bm x}_i= {\bm x}_0\}$, leading to $g^{(n)}(\{ {\bm x}_i= {\bm x}_0\})=1$ for a coherent state.

For a fermionic mode, one has $\braket{ (\hat a^{\dagger})^n \hat a^n}=0$ for $n\geq2$ due to Pauli exclusion principle, leading to the fermionic anti-bunching $g^{(n)}(\{ {\bm x}_i= {\bm x}_0\})=0$ for $n\geq2$.
\newline

The effect of single-mode statistics has been widely probed with ultracold atoms.
In direct analogy with the landmarks experiments of Hanbury Brown and Twiss \cite{hanburybrown:1956, hanburybrown:1956a}, single-atom-resolved observation of bunching and anti-bunching were reported in atomic gases \cite{schellekens:2005,jeltes:2007,cheuk:2016a,tenart:2021a,herce:2023,yao:2025,xiang:2025,dejongh:2025}.
Noise correlation of absorption images was used to unveil underlying spatial ordering, \emph{e.g.} in lattice systems \cite{folling:2005, rom:2006}.
These implementations of the noise correlation technique suffered from the line-of-sight integration inherent to absorption imaging and from a limited resolution in momentum space. 
With detection techniques at the single atom level, perfectly-contrasted bunching was observed in the depletion of a condensate \cite{cayla:2020} and in Mott insulators \cite{carcy:2019a}, while perfectly-contrasted anti-bunching was reported in fermionic clouds \cite{preiss:2019, thomas:2024, dejongh:2025}. 
Similar quantitative analysis of the amplitude of correlations induced by quantum statistics were extended to higher-order correlations, enabling a precise characterization of quantum states through their full counting statistics \cite{dall:2013, carcy:2019a, holten:2021, herce:2023, thomas:2024}.

\subsection{Bosons with contact interactions}

\begin{figure}[!ht]
\includegraphics[width=\columnwidth]{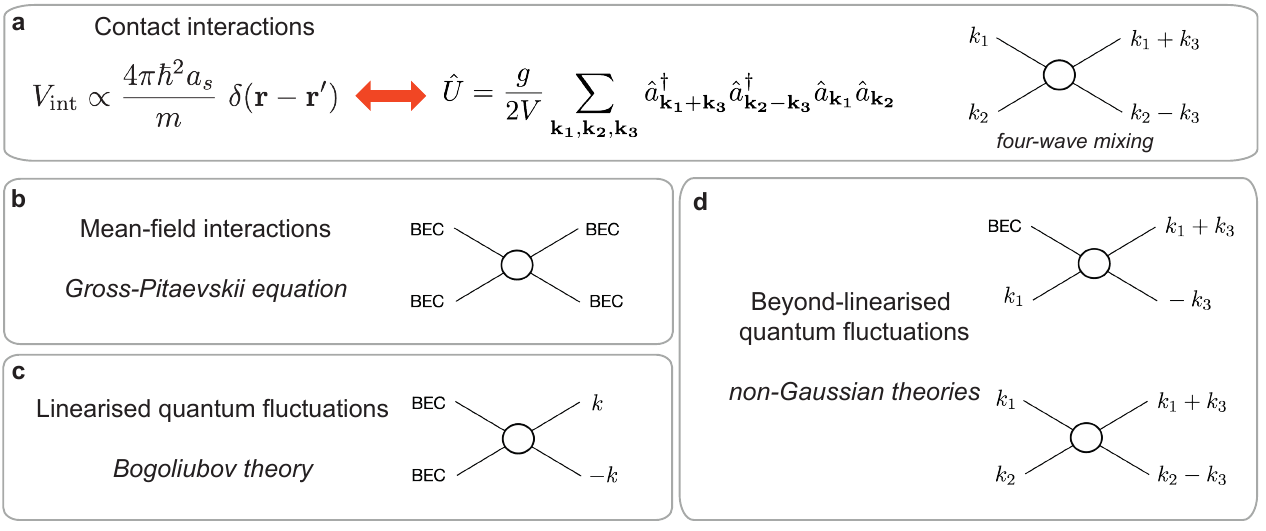}
\caption{(a) Contact interactions and four-wave mixing in momentum space. Here $g=4 \pi \hbar^2 a_s/m$.
(b) Mean-field description of interactions: two-body interactions within the BEC (c) Linearised quantum fluctuations induced by interactions: creation of momentum-correlated pairs exiting the BEC (d) Beyond linearised quantum fluctuations: generation of momentum correlation between 3 or 4 different momentum modes.}
\label{fig3}
\end{figure}

In dilute Bose gases, the atomic density is low enough such that the interaction potential is well approximated by contact interactions, the amplitude of which is set by the two-body $s$-wave scattering length $a_s$. 
The momentum space is a fruitful frame to discuss correlations induced by contact interactions: contact interactions are indeed equivalent to a four-wave mixing term between momentum modes (see illustration in Fig.~\ref{fig3}a). In the following, we consider the Hamiltonian $\hat H=\hat H_0 + \hat U$ with a non-linear (interaction) term $\hat U$ taking the form of a four-wave mixing term, $\hat U\propto\sum_{k_1,k_2,k_3}  \hat a^{\dagger}_{k_1+k_3} \hat a^{\dagger}_{k_2-k_3} \hat a_{k_1} \hat a_{k_2}$. 
Note that such a four-wave mixing term is routinely used to describe experiments in ultracold gases \cite{pitaevskii:2016} as well as in non-linear optics \cite{drummond:2014}. 

Recall that the ground-state of an ensemble of ideal bosons is a Bose-Einstein condensate. 
In the following, the mode of the condensate, {\it i.e.} the ground-state of the non-interacting Hamiltonian $\hat H_0$, corresponds to the mode with a momentum equal to zero, ${\bm k}={\bm 0}$. \footnote{Identifiing the BEC to the mode ${\bm k}={\bm 0}$ is strictly valid only in homogeneous systems with periodic boundary conditions where ${\bm k}$ is a good quantum number. However, it is an excellent approximation in most experimental configurations.} 

\subsubsection{Mean-field approximation -- Gross-Pitaevskii equation}

The simplest way to account for (contact) interactions is to treat them at the mean-field level. This amounts to neglecting quantum fluctuations between different modes and to considering interactions within the single mode of the BEC (Fig.~\ref{fig3}b). The mean-field ground-state remains the Bose-Einstein condensate (BEC), which however differs from the ground-state of ideal bosons through a modification of its wave-function. The mean-field BEC wave-function is described by a non-linear Schrödinger equation, the Gross-Pitaevskii equation, and the condensate fraction ({\it i.e.} the fraction of bosons in the BEC state) is 100\% at zero temperature. 

In terms of cumulants of (momentum) modes, the mean-field approximation implies that any cumulant of order $n>1$ is zero. 
For instance, one has $\braket{ \hat N^2 }_{\rm c}=0$, which implies that $\braket{ \hat N^2 }=\braket{ \hat N }^2$. 
The interaction term, whose sum reduces to the unique contribution of the BEC mode $i=0$, thus takes the form  $\braket{ \hat U }  \propto \braket{ \hat a^{\dagger}_{0} \hat a^{\dagger}_{0} \hat a_{0} \hat a_{0} } = \braket{ \hat N_0 (\hat N_0-1) } =  N_0  ( N_0 -1)$ where $N_0= \braket{\hat N_0}$ is the mean atom number in the BEC mode ${\bm k}={\bm 0}$. Note that the amplitude is simply proportional to the number of interacting atom pairs $N_0  ( N_0 -1)$.

\subsubsection{Linearised quantum fluctuations -- Bogoliubov theory}\label{Subsec:Bogo}

Beyond mean-field descriptions include quantum fluctuations that couple different modes at zero temperature (Fig.~\ref{fig3}c-d).
These quantum fluctuations originate from the non-commutation of the interaction term $\hat U$ with the non-interacting Hamiltonian $\hat H_0$ \cite{pitaevskii:1991}.
They result in the population of modes that are distinct from the BEC, also called the \emph{quantum depletion}.
When quantum fluctuations are a small perturbation of the mean-field solution, they are accounted for by linearising their effect in the Hamiltonian (Fig.~\ref{fig3}c). Systems with linearised quantum fluctuations are referred to as weakly interacting ones. In the case of interacting bosons, the theory with linearised quantum fluctuations is the Bogoliubov theory.

Bogoliubov theory assumes the quantum depletion to be small and approximates the operator of the BEC mode to classical field, $\hat a_0 \sim \sqrt{ N_0}$ where the average number of bosons in the BEC $N_0=\braket{\hat N_0} \sim N$ is close to the total number of particles $N$. The linearisation of quantum fluctuations corresponds to keeping the term $N_0 \sum_{k_1,k_2} [ \hat a^{\dagger}_{k_1} \hat a^{\dagger}_{k_2} + {\rm h.c.} ]$ in the four-wave mixing interactions, in addition to the mean-field contribution.
In terms of cumulants, this approach is equivalent to assuming that cumulants of order $n>2$ are zero.
Inserting the condition $\braket{ \hat a^{\dagger}_{k_1} \hat a^{\dagger}_{k_2} \hat a_{k_3}  }_{\rm c}=\braket{ \hat a^{\dagger}_{k_1} \hat a^{\dagger}_{k_2} \hat a_{k_3} \hat a_{k_4} }_{\rm c}=0$ 
in the interaction terms indeed leads to $\braket{\hat U } \propto N_0 (N_0 -1) + N_0  \sum_{k_1,k_2} \left [ \braket{ \hat a^{\dagger}_{k_1} \hat a^{\dagger}_{k_2} } \braket{ \hat a_{0} \hat a_{0} } + \braket{ \hat a^{\dagger}_{k_1} \hat a^{\dagger}_{0} } \braket{ \hat a_{k_2} \hat a_{0} } + {\rm c.c} \right ]$.
The corresponding interacting Hamiltonian is quadratic in operators and can be diagonalised in a basis of quasi-particles. In other words, the system is well described by a Gaussian theory of non-interacting quasi-particles with quantum fluctuations obeying Gaussian statistics. This is identical to stating that the weakly interacting regime is entirely characterised by connected correlations of order $n \leq 2$.

\begin{figure}[h!]
\includegraphics[width=0.6\columnwidth]{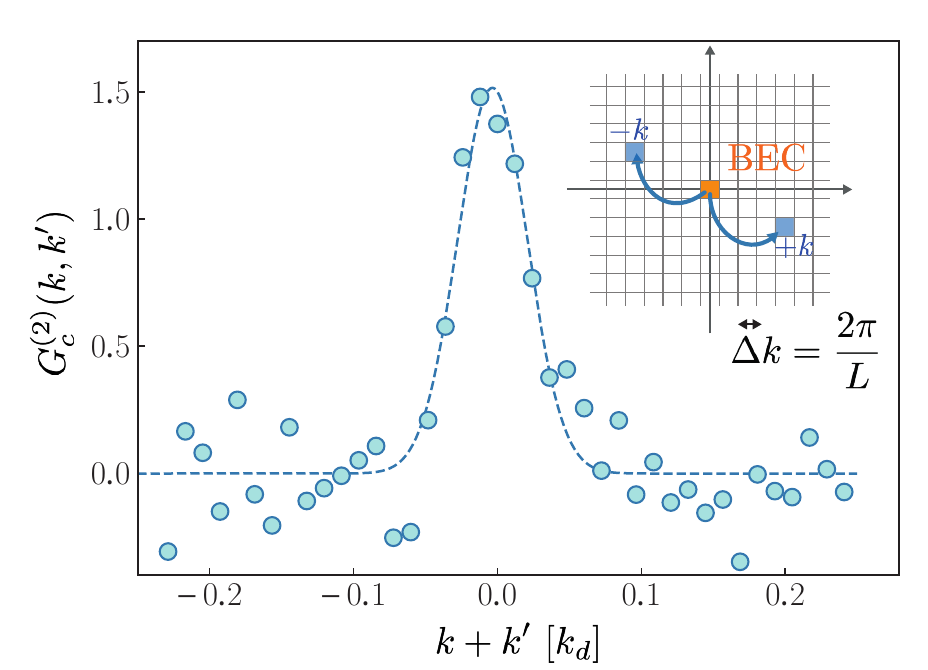}
\caption{Connected correlations $G_{\rm c}^{(2)}({\bm k},{\bm k}')$ in the depletion of a weakly interacting condensate as a function of ${\bm k}+ {\bm k}'$. Inset: illustration of two-mode squeezing induced by interactions. A mode in the momentum space occupies a volume $\Delta k^3$ where $\Delta k = 2 \pi/L$ with $L$ the size of the considered system. The Bose-Einstein condensate (BEC) occupies the mode at ${\bm k}={\bm 0}$ and interactions promote Bogoliubov pairs in modes with opposite momenta $\pm \bm k$. }
\label{fig2}
\end{figure}

Bogoliubov theory is a celebrated example of such a Gaussian theory that treats quantum fluctuations as linear fluctuations. It predicts that the interaction-induced depletion of the BEC takes the form of two-mode squeezed states at opposite momenta (see Fig.~\ref{fig2}). Indeed, momentum conservation in the four-wave mixing process implies that ${\bm k}_2=-{\bm k}_1$ in the contribution $N_0 \sum_{{\bm k}_1,{\bm k}_2} [ \hat a^{\dagger}_{{\bm k}_1} \hat a^{\dagger}_{{\bm k}_2} + {\rm h.c.} ]= N_0 \sum_{{\bm k}_1} [ \hat a^{\dagger}_{{\bm k}_1} \hat a^{\dagger}_{-{\bm k}_1} + {\rm h.c.} ]$. 

In the dilute regime $\sqrt{n a_s^3} \ll1$, Bose gases realise the weakly interacting regime previously described. 
In addition, these platforms are well suited to probe the momentum-correlated pairs predicted by Bogoliubov theory. 
Indeed, single-atom-resolved detection methods after a long free expansion from the trap allow one to access the momentum of individual atoms (see \cite{cayla:2018a} for instance), providing access to the full counting statistics of the atom number $N({\bm k})$ in the mode ${\bm k}$.
The presence of two-mode squeezing at opposite momenta can be revealed from measuring \emph{connected} number correlations at opposite momenta, 
\begin{equation}
    G_{\rm c}^{(2)}({\bm k},-{\bm k})=\braket{ \hat N({\bm k}) \hat N(-{\bm k}) } - \braket{ \hat N({\bm k}) } \braket{ \hat N(-{\bm {\bm k}}) },
\end{equation}
and finding $G_{\rm c}^{(2)}({\bm k},-{\bm k})>0$. In the experiment of Tenart {\it et al.}\cite{tenart:2021a}, this signature was used to reveal Bogoliubov pairing in a gas of weakly interacting bosons, as illustrated in Fig.~\ref{fig2}.
The connected correlations are non-zero only in a narrow range of momenta close to the condition ${\bm k}'=-{\bm k}$, an experimental evidence of the pairing mechanism between modes at opposite momenta. In such experiments, the momentum is a continuous variable which is measured with a resolution $d k$ smaller than the size $\Delta k$ of one mode. The size $\Delta k$ of a mode in momentum space is given by the inverse in-trap size of the gas and it can be measured from the width of the bunching peak in momentum space \cite{cayla:2020}. This is the case in the data reported in Fig.~\ref{fig2}: the size of a mode is given by the width of the correlation peak and corresponds here to roughly $\Delta k \sim 10 dk$. 

\subsubsection{Non-linearised quantum fluctuations -- strongly correlated regime}
\label{Sec-NonGauss-Bosons}

When quantum fluctuations become large enough, their effect cannot be linearised and one must resort to non-Gaussian theories. This is the regime of strongly correlated states. The quasi-particles introduced by linearising the effect of quantum fluctuations acquire a finite lifetime in the strongly correlated regime: they are not eigenstates of the interacting Hamiltonian anymore, which implies that they would interact and decay when excited. 
In other words, a theory of free quasi-particles does not describe the strongly correlated regime.

The non-Gaussian nature of strongly correlated quantum states can in principle be directly revealed from detecting correlated clusters with more than two modes, {\it i.e.} from measuring non-zero connected correlations of order $n>2$ \cite{allemand:2025}. While measuring high-order correlations in a single bosonic mode has long been achieved \cite{dall:2013, herce:2023, thomas:2024}, probing correlations between more than two modes remains scarce \cite{hodgman:2017, carcy:2019a}. 
Alternatively, the measurement of density-density correlations ({\it i.e.} two-body correlations) is sometimes sufficient to obtain signatures of non-Gaussianity, as illustrated recently in two experiments with bosons \cite{ferioli:2024, bureik:2025}. 
In these works, density correlations are used to reveal non-Gaussian statistics in the form of a non-zero fourth-order cumulant of operators. 

\paragraph{Non-zero four-operators cumulants.} 
When one considers an ensemble of bosons with no field coherence, $\braket{ \hat a_i } = \braket{ \hat a_i^{\dagger} } = 0$, the fourth-order cumulant of operators involving two different modes (labeled 1 and 2 respectively) simplifies to $\braket{ \hat a_1^{\dagger} \hat a_2^{\dagger} \hat a_2 \hat a_1  }_{\rm c}
= \braket{ \hat a_1^{\dagger} \hat a_2^{\dagger} \hat a_2 \hat a_1  } 
- |\braket{ \hat a_1^{\dagger} \hat a_2^{\dagger} } |^2
- |\braket{ \hat a_1^{\dagger} \hat a_2 } |^2
- \braket{ \hat N_1 } \braket{ \hat N_2 } $ and leads to 
\begin{equation}
\braket{ \hat N_1 \hat N_2  }_{\rm c} = \braket{ \hat N_1 \hat N_2  } - \braket{ \hat N_1 } \braket{ \hat N_2 } = 
\braket{ \hat a_1^{\dagger} \hat a_2^{\dagger} \hat a_2 \hat a_1  }_{\rm c} + |\braket{ \hat a_1^{\dagger} \hat a_2^{\dagger} } |^2
+ |\braket{ \hat a_1^{\dagger} \hat a_2 } |^2 
\end{equation}
This relation shows that a measurement of negative connected number correlations, $\braket{ \hat N_1 \hat N_2  }_{\rm c}<0$, necessarily corresponds  to $\braket{ \hat a_1^{\dagger} \hat a_2^{\dagger} \hat a_2 \hat a_1  }_{\rm c} < 0$. 
Such a non-zero fourth-order cumulant unambigously indicates that the quantum state is non-Gaussian. 
In the work of Ferioli {\it et al.} \cite{ferioli:2024}, this approach was used to demonstrate the non-Gaussian nature of the light emitted by an ensemble of correlated atomic dipoles (see Section \ref{Section-dissipative-spins}). 
In the experiments of Bureik {\it et al.} \cite{bureik:2025}, it revealed the non-Gaussian nature of momentum correlations in a strongly interacting Bose gas at equilibrium. 
Note that in both the experimental realisations \cite{ferioli:2024,bureik:2025}, the field coherence was shown to be vanishingly small and compatible with being zero. 
This example illustrates that information on non-Gaussian correlators that are not an observable may be obtained from the measured connected correlations, upon minimal assumptions.

\paragraph{Absence of a hierarchy in the four-wave mixing processes at strong interactions.}
As outlined above, there exists a hierarchy in the contributions of the various four-wave mixing processes depicted in Fig.~\ref{fig3} within the weakly interacting regime, where the condensate fraction is close to unity ($N_0 \sim N$). Although not strictly zero, the four-wave mixing processes responsible for generating non-Gaussian states are negligible in this regime. 
In contrast, in the strongly interacting regime, the condensate becomes significantly depleted, and the population of the condensate mode becomes small and comparable to those of the excited modes. As a result, the hierarchical structure of contributions valid in the weakly interacting regime no longer holds and many high-order cumulants have similar contributions.

In the intermediate regime of interactions, where the condensate fraction is appreciably below unity yet remains macroscopic (\emph{e.g.}, $\sim 0.5$), the hierarchy established for weak interactions may still apply. However, the signatures of non-Gaussianity in the many-body state become appreciable as well, as demonstrated in recent experiments at condensate fractions around $\sim 0.75$ \cite{bureik:2025}. This situation shares similarity with the description of the metrological gain in a spin model with a one-axis-twisting term which we discuss in Sec.~\ref{sec:OAT}.

\subsection{Interacting fermions}

Interacting fermions lie at the heart of so-called \emph{strongly correlated materials} whose properties heavily depend on interactions between charge carriers.
Unconventional superconductors (\emph{e.g.} cuprate, or ``high-$T_{\mathrm{C}}$'' superconductors) constitute a paradigmatic example of such materials.
Contrary to conventional superconductors, where phonon-mediated interactions lead to the formation of Cooper pairs \cite{bardeen:1957}, the microscopic origin of the pairing instability in unconventional superconductors is not fully understood, but is closely related to strong electron-electron interactions \cite{morosan:2012}.

Ultracold atom platforms provide a particularly convenient way of studying strongly correlated ensembles of fermions over wide ranges of interactions and densities \cite{ketterle:2008}.
In early experiments using bulk gases, the formation of Cooper pairs was observed through indirect probes related to superfluid behaviour \cite{bourdel:2004, greiner:2003a, jochim:2003, zwierlein:2003, zwierlein:2005, zwierlein:2006}.
The rich phenomenology of the BEC-BCS crossover \cite{randeria:2012}, in which the nature of the pairing correlation changes through a Feshbach resonance, has also been extensively explored with global probes, including RF \cite{chin:2004} and photoemission spectroscopy \cite{stewart:2008}.

In the scope of this article, we rather focus on a few examples of single-atom resolved experimental explorations of correlations emerging in strongly interacting fermionic systems.

\subsubsection{Correlations in bulk ultracold Fermi gases}

Early measurements of correlations in ultracold Fermi gases dealt with density correlations after expansion of the gas.
Superfluidity, for instance, is characterised by a macroscopic population of zero-momentum Cooper pairs.
Such pairing is detected by connected correlations between opposite momenta, $\braket{\hat n(\bm k) \hat n(-\bm k)}_{\mathrm{c}}$ \cite{altman:2004}, similarly to the Bogoliubov pairing of bosons described above.
The formation of local and non-local fermion pairs, in the form of Feshbach molecules, was observed by means of noise correlations in absorption pictures \cite{greiner:2005}.
However, recent experimental efforts led to single-particle detection techniques, allowing to \emph{directly} image such fermionic pairing.
This is illustrated in the following two experiments.

\begin{figure}[h!]
\includegraphics[scale=1]{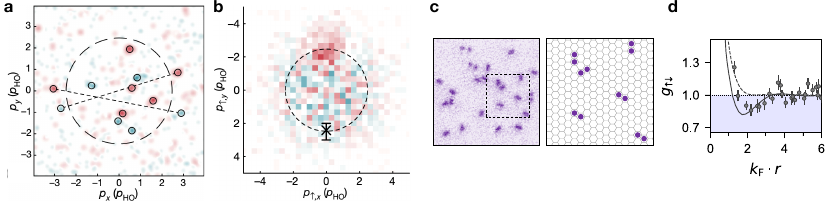}
\caption{
\textbf{a.} Single-atom spin-resolved measurements of a trapped Fermi gas with $N=12$ particles in momentum space.
The dashed circle indicates the Fermi surface.
Momenta are expressed in natural units of the harmonic trap $p_\mathrm{HO} = \sqrt{\hbar m\omega}$.
\textbf{b.} Connected correlations $C^{(2)}(\bm p_\uparrow, \bm p_\downarrow)$, where the momentum of the $\downarrow$-fermion is fixed (cross mark).
\textbf{c.} Single-atom resolved measurements of a dilute, strongly interacting Fermi gas in position space.
\textbf{d.} Normalised correlation function $g_{\uparrow\downarrow}^{(2)}(\bm r)$ as a function of the distance between opposite-spin fermions, showing a strong peak at short distances and anticorrelations at intermediate distances that is not captured by the mean-field BCS theory (dashed line).
The panels \textbf{a} and \textbf{b} were adapted from \cite{holten:2022}, while \textbf{c} and \textbf{d} were adapted from \cite{daix:2025}.
}
\label{figFermions1}
\end{figure}

\paragraph{Identification of individuals BCS pairs.} 
In the experiment of Holten \emph{et al.} \cite{holten:2022}, the ground state of a harmonically trapped Fermi gas with very few particles --- a dozen --- is studied by measuring the momentum of each constituent (see Fig.~\ref{figFermions1}a).
The presence of pairs of fermions lying close to the Fermi surface (dashed circle) is identified by evaluating the connected correlation function $C^{(2)}(\bm p_\uparrow, \bm p_\downarrow) = \braket{\hat n_{\bm p_\uparrow} \hat n_{\bm p_\downarrow}}_{\mathrm{c}}$, represented in Fig.~\ref{figFermions1}b.
There, the cross marks the momentum of the spin-$\downarrow$ particle ; a peak in the correlation function is clearly visible at the opposite side of the Fermi surface, as expected from Cooper pairing.
In their case, the opposite-momentum correlation peak at the Fermi surface is well reproduced by a mean-field BCS theory.

\paragraph{Spatial correlations.}

In a more recent experiment by Daix \emph{et al.} \cite{daix:2025}, a highly dilute two-dimensional Fermi gas is investigated using spatially resolved measurements (Fig.~\ref{figFermions1}c).
As the interaction strength is increased toward the strongly attractive regime, the \emph{normalised} correlation function\footnote{
    Connected and normalised correlations correlation functions are related \emph{via} $g_{\uparrow\downarrow}^{(2)}(\bm r) - 1 = \dfrac{\braket{\hat n_{\bm 0, \uparrow}\hat n_{\bm r, \downarrow}}_{\mathrm{c}}}{\braket{\hat n_{\bm 0, \uparrow}}\braket{\hat n_{\bm r, \downarrow}}}$.
    For spin-balanced homogeneous systems $\braket{\hat n_{\bm r, \sigma}} = n$ and $g_{\uparrow\downarrow}^{(2)}(\bm r) - 1 = \braket{\hat n_{\bm 0, \uparrow}\hat n_{\bm r, \downarrow}}_{\mathrm{c}}/n^{2}$.
}
\begin{equation}
    g_{\uparrow\downarrow}^{(2)}(\bm r) = \frac{\braket{\hat n_{\bm 0, \uparrow}\hat n_{\bm r, \downarrow}}}{\braket{\hat n_{\bm 0, \uparrow}}\braket{\hat n_{\bm r, \downarrow}}}
\end{equation}
develops a pronounced peak as $\bm r \to \bm 0$, indicating the emergence of tightly bound local pairs (Fig.~\ref{figFermions1}d).
This behaviour contrasts with the BCS regime, where pairing involves states of opposite momenta near the Fermi surface, and is therefore non-local in nature.

Interestingly, the experimental data additionally shows that \emph{anticorrelations} develop at intermediate distances $k_{\rm F}r \approx 2$, a feature that is not captured by (mean-field) BCS theory \cite{obeso-jureidini:2022} (Fig.~\ref{figFermions1}d).
The authors attribute this behaviour to non-trivial higher-order correlations.
They evaluate in particular the three point correlations
\begin{equation}
    g_{\sigma\sigma\sigma'}^{(3)}(r) = \frac{\braket{\hat n_{\bm 0, \sigma}\hat n_{\bm r_1, \sigma}\hat n_{\bm r_2, \sigma'}}}{\braket{\hat n_{\bm 0, \sigma}}\braket{\hat n_{\bm r_1, \sigma}}\braket{\hat n_{\bm r_2, \sigma'}}}, \qquad |\bm r_1| = |\bm r_2| = |\bm r_1 - \bm r_2| = r.
\end{equation}
Their experimental setting only allows measuring such correlations in the case where the three particles are equidistant from each other.

\subsubsection{Ultracold fermions in optical lattices}

Once loaded in optical lattices, assuming the lowest energy band is populated, ultracold atoms naturally realise the Hubbard model \cite{jaksch:1998}, originally introduced to describe interacting electrons in solids \cite{hubbard:1963}.
In the case of a spin-$1/2$ system (\emph{e.g.} when considering two hyperfine states of $^{6}\mathrm{Li}$ or $^{40}\mathrm{K}$), the Fermi-Hubbard Hamiltonian writes
\begin{equation}
    \hat H = -t\sum_{\braket{\bm i, \bm j}, \sigma}[\hat c^\dag_{\bm i, \sigma} \hat c_{\bm j, \sigma} + \hat c^\dag_{\bm j, \sigma} \hat c_{\bm i, \sigma}] + U\sum_{\bm i}\hat n_{\bm i, \uparrow}\hat n_{\bm i, \downarrow}.
\end{equation}
Here, $t$ and $U$ are the tunneling amplitudes and on-site interaction energies, $\sigma = \uparrow, \downarrow$ is the spin, $\hat c_{\bm i, \sigma}$ is the annihilation operator for a fermion of spin $\sigma$ on site $\bm i$, and $\braket{\bm i, \bm j}$ designate nearest-neighbour sites.

\paragraph{Antiferromagnetic ordering: two-point correlations}
In typical experiments, the Fermi-Hubbard model is studied on a square lattice\footnote{
    On a bipartite-lattice (such as a square lattice), the Fermi-Hubbard model is particle-hole symmetric.
    Consequently, the phase diagram of the attractive and repulsive cases ($U < 0$ and $U > 0$ respectively) are similar in nature \cite{ho:2009a}.
} in the strongly interacting regime where the physics is dominated by the interactions between fermions ($U \gg t$).
At half-filling --- one fermion per lattice site --- and at sufficiently low temperatures, the system develops short-range antiferromagnetic (AFM) ordering in the repulsive case, which is analogous to a charge density wave (CDW), \emph{i.e.} a staggered checkerboard pattern of empty and doubly occupied sites, in the attractive case.

\begin{figure}[!t]
\includegraphics[scale=1]{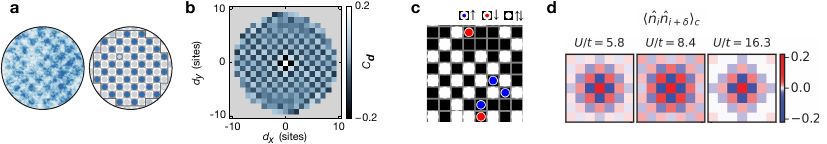}
\caption{
\textbf{a.} Quantum gas microscopy of a repulsive Fermi-Hubbard system close to half-filling, with one spin component removed.
\textbf{b.} Resulting spin correlation map, computed using Eq. \eqref{eq:eqSpinCorrelations}, displaying significant correlations across the entire system.
Panels \textbf{a} and \textbf{b} were extracted and adapted from \cite{mazurenko:2017a}.
\textbf{c.} Quantum gas microscopy of an attractive Fermi-Hubbard system.
\textbf{d.} Resulting density correlations, exhibiting a checkerboard pattern characteristic of CDW ordering, analogous to the AFM ordering of the repulsive case.
Note that this \emph{connected} correlator takes negative values precisely because the disconnected part is subtracted.
Panels \textbf{c} and \textbf{d} were extracted and adapted from \cite{hartke:2023}
}
\label{figSpinSpin}
\end{figure}

Antiferromagnetism can be revealed through the measurement of the spin-spin \emph{connected} correlation function
\begin{equation}
    C_{\mathrm{ss}}^{(2)}(\bm d) = 4\braket{\hat S^z_{\bm 0}\hat S^z_{\bm d}}_{\mathrm{c}},
\label{eq:eqSpinCorrelations}
\end{equation}
where $\hat S^z_{\bm i} = (\hat n_{\bm i, \uparrow} - \hat n_{\bm i, \downarrow})/2$.
In the 2017 Harvard experiment by Mazurenko \emph{et al.} \cite{mazurenko:2017a} (Fig~\ref{figSpinSpin}a,b), a quantum gas microscope, which allowed to locally resolve the atomic density, was used to observe the emergence of extended magnetic ordering --- in practice magnetic correlations extended beyond the system size.

For $U < 0$, fermions with opposite spins tend to pair-up, leading to the formation of doubly occupied sites (doublons).
The symmetries of the attractive model essentially amount to associate empty and doubly occupied sites to the two spin components of the repulsive model ; AFM ordering thus translates to a CDW in the form of a checkerboard pattern, identified through a density correlations $C_{\mathrm{nn}}^{(2)}(\bm d) = \braket{\hat n_{\bm 0} \hat n_{\bm d}}_{\mathrm{c}}$.
First observations of such correlations were reported in \cite{mitra:2018}.
There, correlations are evaluated as a function of local density at $U/t \approx -5$.
More recently, the MIT experiment by Hartke \emph{et al.} \cite{hartke:2023} (Fig.~\ref{figSpinSpin}c,d) which studied the same physics with increasing interaction strength, observed that these correlations peak around $U/t \approx -8$.

In the context of the Fermi-Hubbard model, two point correlations have also been explored in alternative geometries.
In triangular lattices \cite{xu:2023,mongkolkiattichai:2023}, the emergence of magnetic ordering is strongly affected by frustration.
Evidence of pairing between dopants has also been reported in so-called mixed-dimensional geometries \cite{bourgund:2025, hirthe:2023}. 

\paragraph{Highligthing the physics of dopants: three-point spin-charge correlations}
A variety of strongly correlated phases emerge upon doping, \emph{i.e.} when adding or removing fermions \cite{qin:2022}.
Over the past few years, considerable experimental interest has been devoted to the physics of weakly doped antiferromagnets, in a regime associated with the emergence of so-called \emph{magnetic polarons}.
In the framework of condensed-matter physics, polarons are quasiparticles with well-defined excitation spectra, which have been predicted to emerge around mobile dopants in the Fermi-Hubbard model \cite{kane:1989}.
From the point of view of quantum gas microscope experiments, where the position of the spins and the dopants can be measured, these polarons are often described as single dopants ``dressed'' by the background antiferromagnet \cite{grusdt:2018a}.

\begin{figure}[!t]
\includegraphics[scale=1]{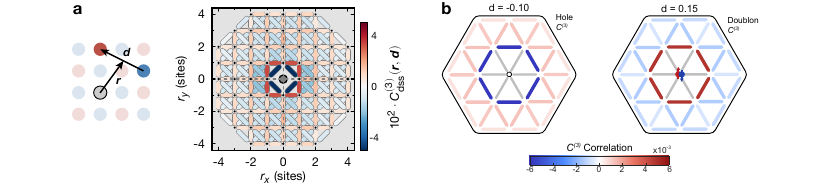}
\caption{
\textbf{a.} Third order dopant-spin-spin correlations measured with a square-lattice Fermi-Hubbard microscope.
In this example (extracted from \cite{chalopin:2024}), significant connected correlations are observed to relatively large distances $|\bm r|$ from the dopant.
Only NN ($|\bm d| = 1$) and NNN ($|\bm d| = \sqrt{2}$) spin-bonds are represented.
\textbf{b.} Third order correlations on a triangular lattice (panel adapted from \cite{prichard:2024}).
Here, the breaking of particle-hole symmetry leads to a strong asymmetry between hole- and doublon-doped systems.
}
\label{figPolarons}
\end{figure}

Third-order spin-charge \emph{connected} correlation functions (Fig.~\ref{figPolarons}a) are well suited to characterise the spatial features of these objects, and have been used in numerous recent experiments on the square \cite{chalopin:2024, koepsell:2019, koepsell:2021a,hartke:2023} and triangular \cite{lebrat:2024,prichard:2024} lattice.
Here, the quantity of interest is
\begin{equation}
    C_{\mathrm{dss}}^{(3)}(\bm r, \bm d) = \frac{4\braket{\hat d_{\bm 0} \hat S^{z}_{\bm r_1} \hat S^{z}_{\bm r_2}}_{\rm c}}{\braket{\hat d_{\bm 0}}}, \quad \text{with} \quad \left\{
    \begin{aligned}
        (\bm r_1 + \bm r_2)/2 & = \bm r \\
        \bm r_2 - \bm r_1 & = \bm d.
    \end{aligned}\right.
    \label{eq:eqPolaron}
\end{equation}
This correlation function quantifies how much the (second-order) spin correlations $\hat S^{z}_{\bm r_1} \hat S^{z}_{\bm r_2}$ are affected by the presence of a dopant $\hat d_{\bm 0}$ nearby.\footnote{
    Here, the dopant takes a different definition depending on whether the system is hole-doped ($\hat d_{\bm i} = (1 - \hat n_{\bm i, \uparrow})(1 - \hat n_{\bm i, \uparrow})$) or doublon-doped ($\hat d_{\bm i} = \hat n_{\bm i, \uparrow}\hat n_{\bm i, \downarrow}$).
}
Remarkably, these \emph{connected} correlations reveal genuine three-body objects composed of a single dopant and two spins, whose existence only relies on two-body interactions.

First usage of these correlation functions was reported in the experiment by Koepsell \emph{et al.} \cite{koepsell:2019}, where the \emph{bare} correlator was used.
Their study was focused on answering the question: what are the spin correlations \emph{conditioned} on the presence of a nearby dopant --- in their case a doubly occupied site.
Their analysis thus consisted in calculating the connected two-point correlation $C_{\mathrm{ss}}^{(2)}(\bm d)$ on post-selected data where the dopant is separated by a distance $\bm r$ to the spin-bond.
This procedure is equivalent to computing the normalised correlator of Eq.~\eqref{eq:eqPolaron}, without removing the disconnected part.
Note that, in a spin-balanced system ($\braket{\hat S^{z}_{\bm i}} = 0$), one finds $C_{\mathrm{dss}}^{(3)}(\bm r, \bm d) = 4\braket{\hat d_{\bm 0} \hat S^{z}_{\bm r_1} \hat S^{z}_{\bm r_2}}/\braket{\hat d_{\bm 0}} - 4\braket{\hat S^{z}_{\bm r_1} \hat S^{z}_{\bm r_2}}$.
In other words, one recovers that the disconnected part of the correlator corresponds to the spin correlation being agnostic to the presence of the dopant.
An example of a connected correlation map is illustrated in Fig.~\ref{figPolarons}a, for NN ($|\bm d| = 1$) and NNN ($|\bm d| = \sqrt{2}$) spin bonds.

More recent studies focused on three-point correlations on the triangular lattice, which break particle-hole symmetry.
There, doublon- and hole-doped systems are completely different.
In the former case, the presence of doublons favors alignment of neighbouring spins and is associated to so-called Nagaoka ferromagnetism, while in the latter case, the presence of holes enhances antiferromagnetism \cite{prichard:2024,lebrat:2024a} (Fig.~\ref{figPolarons}b).
On a triangular lattice, magnetic properties are dominated by the kinetic energy --- or motion --- of the particles, as opposed to the superexchange energy on a bipartite lattice, and enhance considerably the emergence of magnetic polarons \cite{morera:2024}.

\paragraph{Higher-order correlations: strongly correlated regime.}
While second order spin or charge correlations are routinely measured in solid state physics through response functions, third-order correlations that characterise the spatial structure of magnetic polarons are inaccessible in condensed matter devices.
For this reason, higher-order (connected) correlations, which by nature characterise strongly correlated states, have until recently remained elusive to theoretical and experimental condensed matter research.
The power of quantum gas microscopes, and more generally of experiments with single particle detection capabilities, is precisely the ability to measure \emph{connected} correlation functions, in principle up to arbitrary order.
Nevertheless, the measurement and interpretation of correlation functions beyond second or third order in the Fermi-Hubbard model remains mostly exploratory.
The regime where such higher-order correlation functions become significant, in particular, was only reached in recent work --- either by change of the lattice structure \cite{prichard:2024}, or through systematic explorations of doped systems at relatively low temperature \cite{koepsell:2021a, chalopin:2024}.
On the theoretical side, studies that start exploring higher-order correlation functions and their use to characterise strongly correlated states remain very scarce \cite{bohrdt:2021b, muller:2025}.

\section{Spin models}

Spin models also constitute an important class of many-body systems receiving a large attention in the context of quantum simulation and computing, but also for quantum metrology.
The general idea of quantum metrology is to harness quantum many-body effects --- \emph{e.g.} correlations and entanglement --- as a resource for enhancing the sensitivity of measurements.
Intuitively, the regime where entanglement becomes macroscopic corresponds to the strongly correlated regime, in which higher-order cumulants become significant.
In this section, we discuss two different spin systems, the one-axis-twisting (OAT) Hamiltonian relevant in particular for the generation of spin squeezing, and the case of dissipative spins, in which non-trivial correlations due to collective dissipation emerge.
In both cases, the presence of non-Gaussian states can be revealed through a cumulant analysis (see \emph{e.g.} \cite{carollo:2023,fowler-wright:2023}).

\subsection{One-Axis-Twisting (OAT) Hamiltonian}
\label{sec:OAT}
    
The one-axis-twisting Hamiltonian $\hat{H} = \chi \hat{J}_x^2$ is an 
example of a non-linear Hamiltonian useful for metrology \cite{kitagawa:1993,wineland:1994,ma:2011a}. 
It has been implemented in different atomic systems, most often aiming at generating spin-squeezed states with enhanced metrological precision \cite{pezze:2018,riedel:2010,gross:2010,strobel:2014,leroux:2010a,norcia:2018,braverman:2019,pedrozo-penafiel:2020,bohnet:2016,franke:2023}.
Actually, interactions in the OAT model do not generate only spin squeezing in Gaussian states, but they also lead to stronger correlations and non-Gaussian states in certain regimes \cite{evrard:2019}.
Here we use this OAT model to illustrate that cumulants are well suited to capture various regimes of correlations generated by interactions. 
Furthermore, this study also gives the opportunity to discuss the presence, or absence, of a hierarchy in the cumulants. 
We will base our discussion on the experimental results of Ref.\,\cite{evrard:2019}, obtained using Dy atoms in the ground manifold which have spin $J=8$. 
\newline

The dynamics of a large spin under the OAT Hamiltonian generates a Gaussian, squeezed state at early times \cite{kitagawa:1993}. 
Such a squeezed state reduces the uncertainty (or ``squeezes" the noise) in one variable below the standard quantum limit (at the expense of increasing the uncertainty in its conjugate variable). 
Squeezed states have the potential to lead to a metrological gain $\bar G=1/\xi^2$ characterised by the so-called squeezing parameter $\xi$ \cite{wineland:1992, wineland:1994}. 
The squeezing parameter is defined as $\xi^2=2J\,\Delta J_{\rm min}^2/\langle\hat J\rangle^2$, where $\Delta J_{\rm min}$ represents the variance of the spin projected along the axis where it is minimal ({\it i.e.} the ``squeezed" variable). 
Beyond the early dynamics, the OAT creates non-Gaussian states, for instance a cat state $(\ket{J,+J}+\ket{J,-J})/\sqrt2$ at a time $t=\pi/2\chi$ \cite{chalopin:2018}. 
In the case of non-Gaussian states, the sensitivity to an external field can be 
enhanced, even with a large spin variance. 
Generally, the metrological gain is not defined by the squeezing parameter, but rather by the Hellinger distance that measures the similarity between two probability distributions \cite{strobel:2014,evrard:2019}.

\begin{figure*}[h!]
\includegraphics[scale=1]{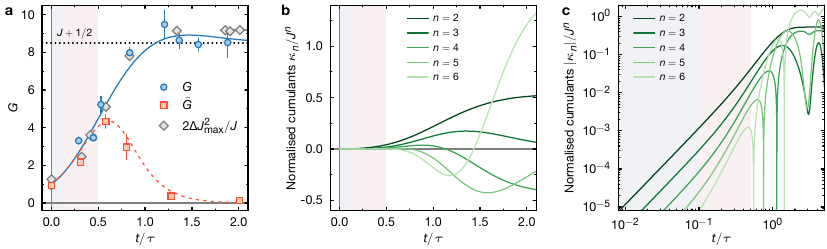}
\caption{
{\bf a}. Dynamics of the squeezing (Gaussian) metrological gain $\bar G$ and of the generalised (non-Gaussian) gain $G$, defined from the Hellinger distance. 
$\tau=(\sqrt{2J}\chi)^{-1}$ is the time scale of the OAT dynamics. 
The solid and dashed lines are theoretical predictions. 
The markers are measurements reported in \cite{evrard:2019}. 
{\bf b}. Normalised cumulants of the operator $\hat J_z$ as a function of time. 
The $n^{\mathrm{th}}$-order cumulant $\kappa_n \equiv \braket{\hat J_z^n}_{\rm c}$ is normalised by the $n^{\rm th}$ power of the total spin length, $J^n$.
{\bf c}. Same plot as in {\bf b} in log-log scale.
In all the panels, the blue shaded area corresponds to early times $0\leq t/\tau\leq 0.1$ for which the hierarchy of cumulants is well-established and stationary. 
The red shaded area corresponds to intermediate times $0.1 \leq t/\tau\leq 0.5$ where the hierarchy, although still valid, evolves significantly. 
At later times $t/\tau\geq 0.5$, no hierarchy can be identified.
}
\label{fig3:OAT}
\end{figure*}

The difference between the regimes where one can assume a Gaussian state and use the squeezing parameter to evaluate the metrological gain, and the one where this assumption breaks down is highlighted in Fig.\,\ref{fig3:OAT}a adapted from \cite{evrard:2019}. 
Both the metrological gain given by the squeezing parameter ($\bar G$) and the generalised one ($G$) are plotted as a function of duration $t$ of the OAT dynamics. 
While their values coincide at early times, $t/\tau \leq 0.5$, the squeezing parameter is unable to capture the metrological gain associated with the regime of non-Gaussian states. 

We now discuss the OAT dynamics through the time evolution of the cumulants of the operator $\hat J_z$, rather than the metrological gain. Those are plotted in  Fig.\,\ref{fig3:OAT}b-c.
For very small evolution durations, $t/\tau \leq 0.1$ (blue regions in Fig.~\ref{fig3:OAT}), a well-defined hierarchy in the cumulants is found, with $\kappa_{n+1} \ll \kappa_n$ for $n\geq2$ (see Fig.~\ref{fig3:OAT}c). 
In this regime, neglecting the contributions from cumulants of order larger than $n=2$ is an excellent approximation. 
This is the regime where the Gaussian approximation holds, accounting only for linearised fluctuations and considering Gaussian states (see \ref{Subsec:Bogo}). 
As expected, the metrological gain is given by squeezing $G \simeq \bar G$ in this regime. 
Although negligible at these short dynamics, high-order cumulants always take finite values.
At intermediate durations, $0.1 \leq t/\tau \leq 0.5$ (red region in Fig.~\ref{fig3:OAT}), the hierarchy in the cumulants is conserved. 
However, the amplitudes of high-order cumulants $n>2$ approach that of the second-order cumulant, with the consequence that neglecting the contributions from high-order cumulants is not a good approximation anymore. 
This implies that quantum states are not well approximated with Gaussian states and that the metrological gain $G$ starts departing from that given by the squeezing parameter $\bar G$ (see Fig.~\ref{fig3:OAT}a).  
At later times, $0.5 \leq t/\tau$, the hierarchy of the cumulants breaks down, with high-order ones $n\geq3$ contributing similarly to the $n=2$ cumulant. 
The gains $G$ and $\bar G$ strongly differ as well. 
This regime corresponds to a strongly correlated regime where non-Gaussian correlations play a central role in the physical properties. 
Importantly, we find that, similarly to the expectations for bosons \cite{robinson:1965}, no hierarchy in the cumulants exists in the strongly correlated regime for the considered system of interacting spins. 

\subsection{Dissipative spin models}
\label{Section-dissipative-spins}

As a last example, we consider a spin system with collective dissipation.
There, one is interested in the emergence of correlations as a result of the collective dissipation. One situation where this could occur
is Dicke superradiance \cite{dicke:1954}, where one considers an ensemble of identical two-level atoms with permutational symmetry.
In this case, starting from a fully inverted system, Dicke showed that correlations of the type $\braket{\hat \sigma^+_i\hat \sigma^-_j}$ emerge due to collective spontaneous emission while the individual dipole of each atom remains zero, {\it i.e.} $\braket{\hat \sigma^-_i}=0$ \cite{gross:1982}.
More generally, the question of how correlations form in a dissipative spin system when collective dissipation competes with an external driving field is the topic of intense investigation \cite{lee:2013,olmos:2014,parmee:2018,parmee:2020,henriet:2019a,verstraelen:2023,stitely:2023,rubies-bigorda:2025}. 
\newline

We will focus here on one type of dissipative spin system: two-level atoms coupled to each other via their dipole radiation and collectively dissipating by spontaneous emission.
The usual treatment of this problem traces out the field degrees of freedom and results in a master equation for the two-level atoms' density matrix \cite{lehmberg:1970}.
Solving the full master equation to describe the dynamics is out of reach.
As a consequence, one resorts to approximations to describe the system's dynamics and/or steady state.
One such approximation is to perform a cumulant expansion, truncated at a given order \cite{kirton:2019,robicheaux:2021,plankensteiner:2022,rubies-bigorda:2023,verstraelen:2023}. 
In particular, it is known that the superradiant pulse, even in large ensembles outside of the Dicke regime can be well described by mean-field dynamics, with fluctuations playing a role mostly at the beginning of the pulse \cite{gross:1982}, which can be captured by a second-order (Gaussian) truncation.
Increasing the truncation order is rapidly very costly numerically, and it remains a challenge to perform beyond Gaussian simulations despite recent progress, for instance considering systems with spatial symmetries \cite{holzinger:2025}.
In addition, it is known that a cumulant expansion might fail \cite{fowler-wright:2023}, for instance to describe some observables in specific regimes such as the density matrix of the system during subradiance due to the presence of high-order correlations \cite{henriet:2019a}.
Another theoretical method, which does not truncate the cumulant expansion but somehow assumes that correlations do not grow strongly during the dynamics is the truncated Wigner approximation, which has been adapted for spin models recently \cite{mink:2023,hosseinabadi:2025}, and has been successful at describing the dynamics of atoms radiating collectively via a nano-fiber waveguide \cite{tebbenjohanns:2024,bach:2024}.
From the experimental point of view, measuring cumulants is useful if one wants to flag the appearance of correlations in the ensembles.
As we will see below it is also a natural byproduct of measurements of the statistics of the radiated light. 

To relate collective light scattering to correlations in atomic ensembles, we write the expression of the electric field emitted by the ensemble in terms of atomic operators: 
\begin{equation}
    \hat E^+(\bm{R}) = \sum_n G(\bm R-\bm r_n)\hat \sigma^-_n,
\end{equation}
where $G(\bm R-\bm r_n)$ is the Green's function for propagation of the electric field \cite{asenjo-garcia:2017}. 
From this one obtains the intensity $I=\braket{ \hat E^-\hat E^+}$ emitted by $N$ atoms in a direction $\bm e_k$: 
\begin{equation}
    I_N(\bm e_k)=I_1(\bm e_k)\sum_{m,n}e^{-i\bm{k}\cdot(\bm r_n-\bm r_m)}\braket{\hat \sigma^+_n\hat \sigma^-_m}.
    \label{Eq:intensity}
\end{equation}
Here, $\bm{k}=k_0\bm e_k$, $k_0=2\pi/\lambda_0$ is the wavevector corresponding to the atomic transition and $I_1(\bm e_k)$ is the intensity emitted by a single isolated atom. 
Equation \eqref{Eq:intensity} shows that the presence of two-atom correlations with the right phase factor can enhance the intensity emanating from an atomic ensemble in a particular direction. 
This is how Dicke superradiance results in a stronger emission than that of independent atoms. 
The observation of a superradiant burst of light \cite{skribanowitz:1973,gross:1976,gibbs:1977,ferioli:2021,liedl:2024} from an initially inverted cloud of atoms (\emph{i.e.} in a product state where all atoms are in the excited state $\ket{e}$) indicates the spontaneous appearance of two-atom correlations. 
Since during the decay $\braket{\hat \sigma_n^+} = 0$, it directly reveals a non-zero second-order cumulant $\braket{\hat \sigma^+_n\hat \sigma^-_m}\neq\braket{\hat \sigma^+_n}\braket{\hat \sigma^-_m}$.

\begin{figure*}[h!]
\includegraphics[width=0.5\textwidth]{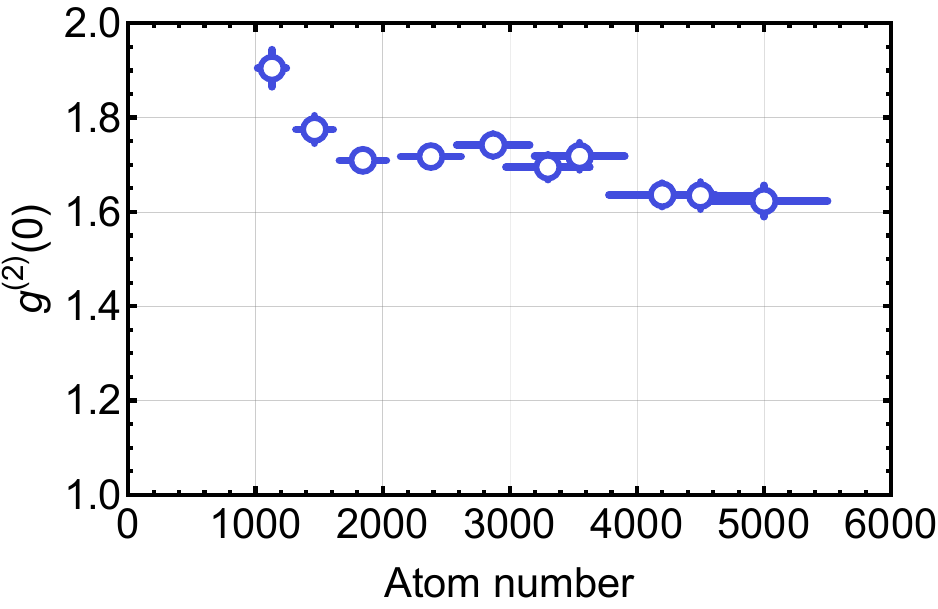}
\caption{Measurement of the second order coherence  at equal time ($g^{(2)}(0)$) of light emanating from an elongated cloud of cold Rb atoms continuously driven by a resonant laser beam, in steady-state. The data is consistently violating the Siegert relation since $g^{(2)}(0)<2$, which marks non-Gaussian correlations. Data reprinted from Ref.\,\cite{ferioli:2024}.}
\label{fig:g2_light}
\end{figure*}

At this level, the state of the light field radiated by the ensemble may still be described as a Gaussian field.
However, preparing quantum non-Gaussian field states is an 
outstanding goal of quantum optics \cite{lvovsky:2020,walschaers:2021,lachman:2022,olmos:2020}, 
motivating the investigation of correlations beyond second order.
The natural experimental observable to look at is thus the second-order correlation function $g^{(2)}(\delta t)$ \cite{loudon:2000,steck:2007}.
Its measurement during a superradiant pulse was performed recently, both in free-space atomic ensembles \cite{ferioli:2025} and with atoms coupled to a nanofiber \cite{bach:2024}: there, the transition from $g^{(2)}(0)\simeq2$ to $g^{(2)}(0)\simeq1 $ revealed that superradiance establishes correlations between the emitting dipoles.
Finally, non-Gaussian correlations have been observed in the light radiated by a free-space, driven atomic cloud, \emph{in steady-state} \cite{ferioli:2024}.
In that case, the measurements of $g^{(2)}(\delta t)$ revealed a violation of the Siegert relation (Eq.~\eqref{Eq:thermal}), visible in Fig.~\ref{fig:g2_light}, where one observes that $g^{(2)}(0)<2$.
As discussed above, this indicates either (i) a non-zero field coherence $\braket{ \hat E^-}$ or (ii) a non-zero fourth order cumulant.
Measurements of $g^{(1)}$ revealed a negligible field coherence.
As a consequence that experiment showed the existence of a non-zero fourth order cumulant.
This means that one needs to account at least for cumulants to the fourth order to describe this system, which has so far prevented a theoretical prediction.
We note that to be a useful source of non-Gaussian light for quantum optics, one also has to demonstrate its ``quantumness'', \emph{i.e.}~Wigner negativity, which was not the case in the experiments of ref.~\cite{ferioli:2024}.

\section{Conclusion}

This article covered a few recent examples of ultracold atoms experiments measuring connected correlations in quantum many-body systems of bosons, fermions and of spins. Through these examples, it illustrates how microscopic connected correlations are  well suited to characterizing quantum many-body states in the strongly correlated regime. More precisely, they unveil non-Gaussian features --- those that go beyond the predictions of free quasi-particle theories --- and they reveal the presence of correlated clusters involving more than two particles or modes. 
The ability to measure such high-order connected correlations is a direct consequence of the single-particle detection capabilities of modern experiments using synthetic quantum systems. As such, the ideas presented in this article apply to other platforms as well, including trapped ions \cite{leibfried:2005, bohnet:2016}, Rydberg atoms \cite{browaeys:2020}, superconducting qubits \cite{mi:2024} and photonic systems \cite{lvovsky:2020, scarpelli:2024}.

Beyond the identification of non-Gaussianity and correlated clusters discussed here, strongly correlated states are also often expected to exhibit entanglement. 
However, characterizing macroscopic entanglement properties in quantum many-body systems remains a significant challenge. 
In connection to this article, we find it interesting to ask whether the measurement of connected correlations could also prove useful in this context. 
For instance, in the case of Gaussian states, such as squeezed states, connected correlations have proven to be valuable tools \cite{bergschneider:2019, gondret:2025}.
Nevertheless, multipartite-entangled states typically exhibit non-Gaussian features \cite{pezze:2018, strobel:2014}, and the extent to which connected correlations can meaningfully capture entanglement in the strongly correlated regime remains an open and active question.

\section{Acknowledgements}
We acknowledge stimulating and insightful discussions with Maxime Allemand, Alain Aspect, Denis Boiron and the \emph{Quantum Gases} group at Institut d'Optique. 
We thank Tarik Yefsah for providing the figures of Fig.~\ref{figFermions1}c,d.
This work was supported by the France 2030 program (ANR-22-PETQ-0004, project QuBitAF) the Region Ile-de-France in the framework of the DIM QuanTiP, the European Research Council (Advanced grant No. 101018511-ATARAXIA, ERC StG CORSAIR,
101039361), and 
the Horizon Europe programme HORIZON-CL4- 2022-QUANTUM-02-SGA project 101113690 (PASQuanS2.1).

\bibliographystyle{crunsrt}

\def\bysame{\leavevmode ---------\thinspace}
\makeatletter\if@francais\providecommand{\og}{<<~}\providecommand{\fg}{~>>}
\else\gdef\og{``}\gdef\fg{''}\fi\makeatother
\def\cdrandname{\&}
\providecommand\cdrnumero{no.~}
\providecommand{\cdredsname}{eds.}
\providecommand{\cdredname}{ed.}
\providecommand{\cdrchapname}{chap.}
\providecommand{\cdrmastersthesisname}{Memoir}
\providecommand{\cdrphdthesisname}{PhD Thesis}

\end{document}